\documentclass{optica-article}

\journal{opticajournal} 
\articletype{Research Article}
\usepackage{multirow,booktabs}
\usepackage{tabulary}
\usepackage{graphicx}
\usepackage{hyperref}
\hypersetup{
    colorlinks=true,
    linkcolor=green,
    filecolor=green,
    urlcolor=blue,
    citecolor=blue
}
\usepackage{lineno}
\usepackage{amsmath}
\usepackage[T1]{fontenc}
\usepackage[utf8]{inputenc}

\begin{document}

\title{The Mechanism of Chirality in Plasmonic Structures: Asymmetric Resonance Modes}

\author{Abduweli Abdukerim}
\address{College of Physics and Electrical Engineering, Kashi University, Kashi 844000, China}

\email{\authormark{$\ast$}abduwelee@snnu.edu.cn, aasci@ksu.edu.cn}
.

\begin{abstract*}
This study presents a theoretical framework for understanding the mechanism of chirality generation and circular dichroism (CD) in chiral nanostructures, with a particular focus on the role of asymmetric plasmon resonance mode distributions. Although previous studies have linked chiroptical properties to resonance differences, the specific nature and origin of these differences have not been fully elucidated. Here, we provide a more detailed explanation by quantitatively analyzing the asymmetric electric-field components arising from mode asymmetry. Specifically, we examine the near-field symmetry of individual nanorods and corner-stacked coupled structures (CSCS). The results demonstrate that chirality arises from asymmetric electric-field distributions, which give rise to distinct phase and amplitude responses of the electromagnetic fields in the composite structure under right-handed circularly polarized (RCP) and left-handed circularly polarized (LCP) illumination. The absorption and CD spectra, which exhibit distinct features around $910\,\mathrm{nm}$, support this mechanism and provide valuable insights into the design of chiroptical materials. The proposed mechanism is not restricted to a specific chiral structure or resonance and offers a broader explanation for the generation of circular dichroism.
\end{abstract*}

\section{Introduction}

Chirality is a property whereby an object or molecule cannot be superimposed on its mirror image \cite{1,2}. This intriguing feature has attracted considerable attention because of its widespread occurrence in nature and its significant implications across a variety of scientific disciplines, such as chemistry, biology, and physics \cite{3,4,5}. Chiral nanostructures exhibit remarkable phenomena, including circular dichroism (CD) and asymmetric transmission (AT) \cite{MA2026133598,10.1021/acs.nanolett.5c02945,10.1021/acs.nanolett.5c05530}. These phenomena result in observable differences in light absorption and transmission when the structures interact with right-handed circularly polarized (RCP) and left-handed circularly polarized (LCP) light. CD spectroscopy has become an indispensable tool for studying chirality in various materials, including chiral biomolecules such as proteins, DNA, and RNA \cite{11.0}. It has found broad applications in diverse areas, such as materials science, biochemistry, and pharmaceuticals \cite{6,7,8.0}. In parallel, the AT effect offers considerable potential for polarization control \cite{9.0}, as well as for acoustic noise control \cite{10}. Furthermore, advances in fabrication and characterization techniques have greatly broadened the applications of chiral plasmonic structures. In particular, such structures have attracted significant attention owing to their potential in diverse fields, including chiral sensing \cite{11.0,12.0,8,14.0}, optical forces \cite{15.0,16.0,17.0}, polarization photodetectors \cite{18.0,19.0,20.0}, and chiral catalysis \cite{21.0,22.0}. These advances have created new opportunities for manipulating light and investigating complex molecular interactions.

The unique properties of these nanostructures arise from the distinctive surface plasmon resonance (SPR) modes supported on their surfaces. A comprehensive understanding of the associated localized electromagnetic fields is crucial for elucidating these properties and further optimizing their performance. However, the spatial distribution of the localized electromagnetic fields is complex, with variations in field strength and polarization at each point. In addition, their quantitative characterization is constrained by the computational capabilities of simulation software. Therefore, most studies remain limited to qualitative analyses \cite{8}. These limitations also extend to investigations of chiral effects. Most existing studies on plasmonic chirality remain largely qualitative and lack a clear physical interpretation of the mechanisms underlying chirality \cite{3c18947,4c06157,Zhao:21,Qiao:24}. Among the available theoretical approaches, the Born--Kuhn model has been widely accepted and frequently employed as a classical framework for interpreting chiral behavior, in which chiral structures are typically represented as assemblies of coupled resonators \cite{26.0,27.0,28.0}. Under excitation by RCP and LCP light, chiral materials generate different resonant modes and exhibit varying degrees of coupling. However, the underlying reasons for these differences remain unclear, and no definitive conclusion has yet been reached. In this context, recent studies have made significant progress by calculating the evolution of surface plasmon modes and their oscillatory behavior over an entire cycle \cite{Abdukerim_2024,Abdukerim:24}. Such advances in computational methods, characterization techniques, and physical insight have enabled more systematic investigations of near-field interactions on nanostructure surfaces. Building on these developments, it has become possible to further uncover the mechanisms underlying the formation of distinct resonant modes and their roles in the emergence of plasmonic chirality.

In this paper, we introduce a corner-stacked coupled structure (CSCS) and use it to investigate the mechanisms underlying the generation of different resonant modes within the framework of the Born--Kuhn model. Our findings demonstrate that surface plasmon resonance generates a near field that is asymmetric with respect to the symmetry axis around an achiral structure under circularly polarized illumination. This asymmetry leads to different coupling behaviors of the achiral components within the chiral assembly under RCP and LCP illumination, resulting in distinct optical responses of the chiral structure. These insights contribute to a deeper understanding of chiral plasmonics and are expected to provide valuable guidance for future research in this area.

\section{Structure and Computational Method}

\begin{figure}[b]
\centering
\includegraphics[width=1\textwidth]{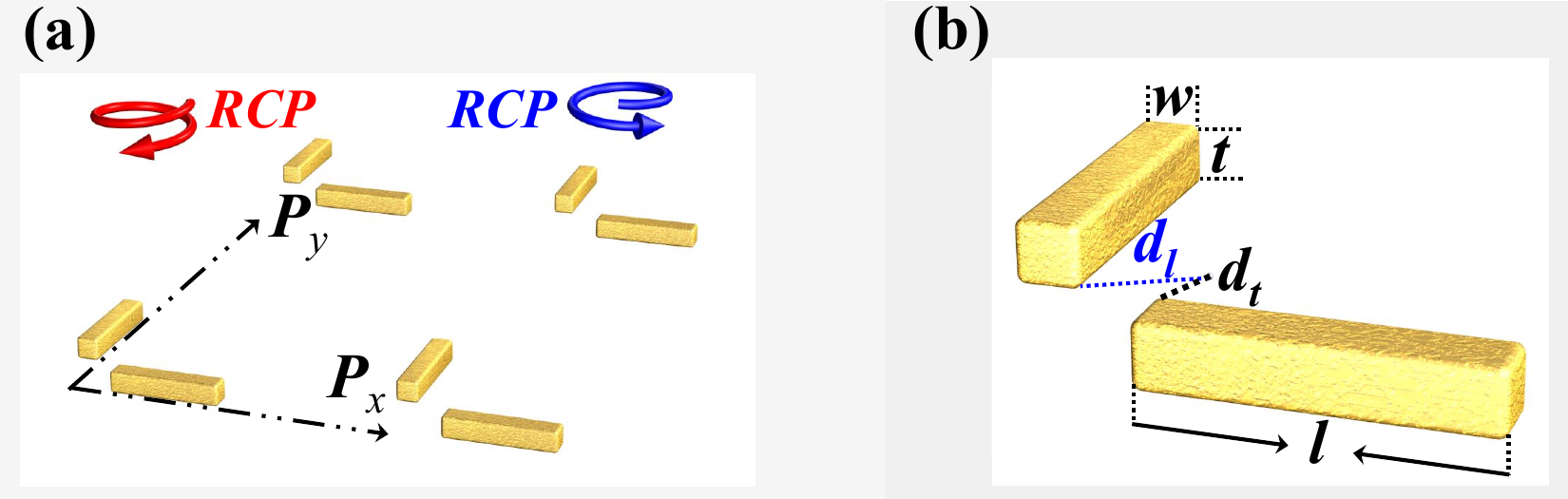}
\caption{Schematic of the corner-stacked nanorod structure: (a) array and (b) unit cell. The width and height are $w=h=40\,\mathrm{nm}$, and the length of the rods is $l_s=200\,\mathrm{nm}$. The distance between the coordinate axis and the nanorods is $d=300\,\mathrm{nm}$.}
\label{fig1}
\end{figure}

Figure~\ref{fig1}(a) illustrates a schematic of our proposed structure, a corner-stacked configuration consisting of two identical, mutually perpendicular nanorods. The origin of chirality can be identified by comparing the SPR characteristics of the coupled structure with those of each individual nanorod before coupling. The simulations were performed using the finite element method. The height ($h$) and width ($w$) of the nanorods are identical and were both set to $40\,\mathrm{nm}$. The length of the nanorods is $l_s=200\,\mathrm{nm}$. For simplicity and computational accuracy, we used full-field calculations instead of scattered-field calculations. The periods of the structure were set sufficiently large to neglect interactions between adjacent unit cells, with $P_x=P_y=4000\,\mathrm{nm}$. Periodic boundary conditions were applied in both the $x$ and $y$ directions.

The CSCS was composed of gold, and its refractive index was calculated using the Brendel--Bormann model \cite{29.0}. The entire structure was illuminated by circularly polarized light (CPL) propagating along the negative $z$ direction. The light was introduced through an excitation port with an input power of $1\,\mathrm{W}$. Perfectly matched layers were placed around the structure to absorb outgoing light and minimize reflections.

The absorption was calculated by integrating the resistive losses over the entire metal volume. $A_R$ denotes the absorption under RCP illumination, and $A_L$ denotes the absorption under LCP illumination. CD is defined as $\mathrm{CD}=A_R-A_L$. The absorption is calculated as
\begin{equation}
    A=
    \frac{1}{2P_{\mathrm{inc}}}
    \int_{V_{\mathrm{m}}}
    \operatorname{Re}\left(\mathbf{J}^{\ast}\cdot\mathbf{E}\right)
    \,\mathrm{d}V,
\end{equation}
where $V_{\mathrm{m}}$ is the metal volume, $\mathbf{J}$ is the induced volume current density, $\mathbf{E}$ is the electric field, and $P_{\mathrm{inc}}$ is the incident power.

\section{Results and Discussion}

\begin{figure}[h]
\centering
\includegraphics[width=1\textwidth]{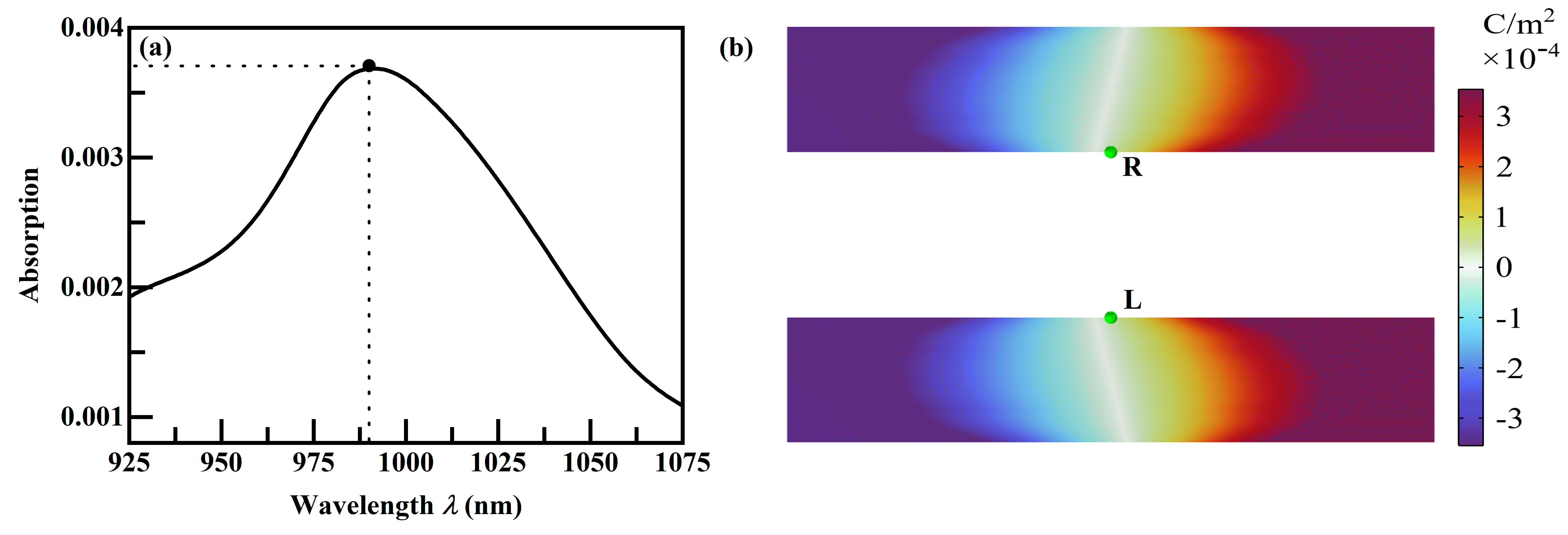}
\caption{(a) Absorption spectrum of an individual nanorod. (b) Surface charge distributions of the nanorod under RCP and LCP illumination at $990\,\mathrm{nm}$.}
\label{figabs}
\end{figure}

Before calculating the optical properties of the coupled nanorods, we first examined the optical properties of each individual nanorod. This step is crucial because the chiral SPR behavior of the coupled nanorods can be better understood through comparison with their individual counterparts. Figure~\ref{figabs}(a) presents the absorption spectrum of an individual nanorod. A peak is observed at $990\,\mathrm{nm}$. To identify the SPR mode, we analyzed the surface charge distributions of the nanorod under RCP and LCP illumination. Figure~\ref{figabs}(b) presents the surface charge distributions, from which it is clear that the mode originates from a dipolar resonance along the long axis of the nanorod. The charge distributions are asymmetric with respect to the symmetry axis of the structure. Nevertheless, the distributions under RCP and LCP illumination are mirror-symmetric counterparts of each other. This feature is not confined to a particular phase; for any given surface charge distribution under one circular polarization, a corresponding mirror-symmetric distribution can always be found under illumination with the opposite handedness. As a result, the electromagnetic oscillations at the mirror-symmetric points are symmetry-equivalent.

\begin{table}[t]
\caption{Amplitudes and initial phases of the $x$, $y$, and $z$ components of the electric field under RCP and LCP excitation.}
\label{tab:1}
\centering
\begin{tabular}{ccc}
\toprule
Component & RCP & LCP \\
\midrule
$x$ component &
\begin{tabular}{@{}c@{}}
$E_{0x}=1.82118\times 10^7\,\mathrm{V/m}$ \\
$\phi_{0x}=-139.90^\circ$
\end{tabular} &
\begin{tabular}{@{}c@{}}
$E_{0x}=1.82065\times 10^7\,\mathrm{V/m}$ \\
$\phi_{0x}=-139.91^\circ$
\end{tabular} \\
$y$ component &
\begin{tabular}{@{}c@{}}
$E_{0y}=0.98796\times 10^6\,\mathrm{V/m}$ \\
$\phi_{0y}=-11.83^\circ$
\end{tabular} &
\begin{tabular}{@{}c@{}}
$E_{0y}=1.02019\times 10^6\,\mathrm{V/m}$ \\
$\phi_{0y}=168.88^\circ$
\end{tabular} \\
$z$ component &
\begin{tabular}{@{}c@{}}
$E_{0z}=6.09954\times 10^6\,\mathrm{V/m}$ \\
$\phi_{0z}=-4.19^\circ$
\end{tabular} &
\begin{tabular}{@{}c@{}}
$E_{0z}=6.10676\times 10^6\,\mathrm{V/m}$ \\
$\phi_{0z}=-4.20^\circ$
\end{tabular} \\
\bottomrule
\end{tabular}
\end{table}

\begin{figure}[h]
\centering
\includegraphics[width=1\textwidth]{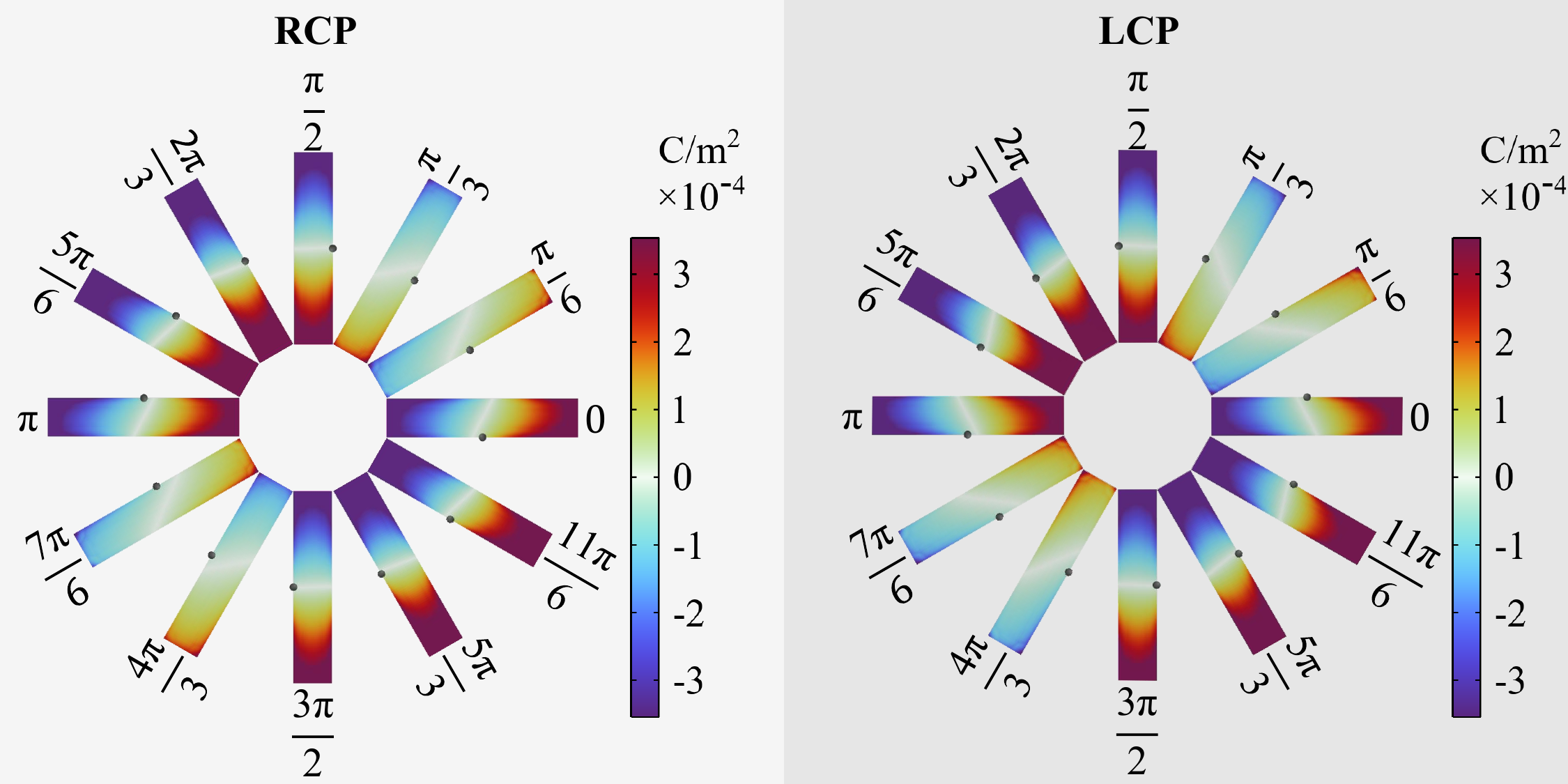}
\caption{Surface charge distributions of an individual nanorod at different phases over one oscillation cycle under RCP and LCP illumination at $990\,\mathrm{nm}$.}
\label{mircha}
\end{figure}

To further describe the SPR behavior quantitatively, we derived the electric-field oscillation equations at a wavelength of $990\,\mathrm{nm}$ by investigating the full-cycle oscillation of the surface charge under both RCP and LCP excitation \cite{Abdukerim_2024,Abdukerim:24}. The electric-field oscillation can be expressed as
\begin{equation}
    E=E_0\sin(\omega t+\phi_0),
\end{equation}
where $\omega=2\pi c/(990\,\mathrm{nm})$. The amplitudes $E_0$ and phases $\phi_0$ at the mirror-symmetric points \textbf{R} and \textbf{L} are summarized in Table~\ref{tab:1}. These points are marked in Fig.~\ref{figabs}(b). The $x$ and $z$ components at the two points have nearly identical amplitudes and initial phases. The $y$ components have nearly identical amplitudes but an approximately $180^\circ$ phase difference, which arises from the phase relation of the $y$ component of the incident RCP and LCP fields. Thus, the electric-field oscillations at the two points are symmetry-equivalent. To show the mirror symmetry of points \textbf{R} and \textbf{L} more clearly, Fig.~\ref{mircha} presents the charge distributions at different phases. Although the charge distribution evolves over time, for any phase under RCP illumination, a mirror-related charge distribution can be found under LCP illumination.

\begin{figure}[h]
\centering
\includegraphics[width=1\textwidth]{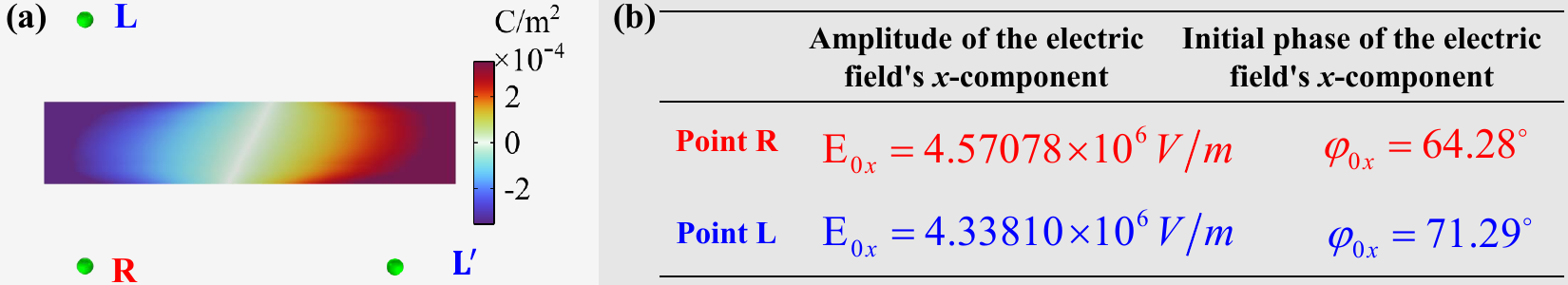}
\caption{(a) Surface charge distribution of an individual nanorod, with points \textbf{R}, \textbf{L}, and $\mathbf{L}^{\prime}$ indicated. (b) Amplitudes and initial phases of the $x$ component of the electric field at points \textbf{R} and \textbf{L}.}
\label{ISO}
\end{figure}

For isolated transverse and longitudinal nanorods, the oscillations are equivalent after accounting for their orthogonal orientations. However, the oscillation in the $x$ direction of the transverse nanorod corresponds to the oscillation in the $y$ direction of the longitudinal nanorod, and the oscillation in the $y$ direction of the transverse nanorod corresponds to the oscillation in the $x$ direction of the longitudinal nanorod, resulting in a $90^\circ$ phase difference. Note that the electromagnetic near field of a single nanorod is not mirror symmetric with respect to its symmetry axis; instead, it exhibits central symmetry. Figure~\ref{ISO}(a) illustrates this behavior in our calculations. Points \textbf{R} and \textbf{L} are mirror-symmetric points, whereas points \textbf{L} and $\mathbf{L}^{\prime}$ are centrally symmetric points around the nanorod. Although points \textbf{R} and \textbf{L} are mirror symmetric, the electric fields at the two points are different. The amplitude and initial phase of the $x$ component at point \textbf{R} are $E_{0x}=4.57078\times 10^6\,\mathrm{V/m}$ and $\phi_0=64.28^\circ$, respectively, whereas the corresponding values at point \textbf{L} are $E_{0x}=4.33810\times 10^6\,\mathrm{V/m}$ and $\phi_0=71.29^\circ$. However, the oscillations at points \textbf{L} and $\mathbf{L}^{\prime}$ have the same amplitude and waveform, with a $180^\circ$ phase difference.

\begin{figure}[h]
\centering
\includegraphics[width=1\textwidth]{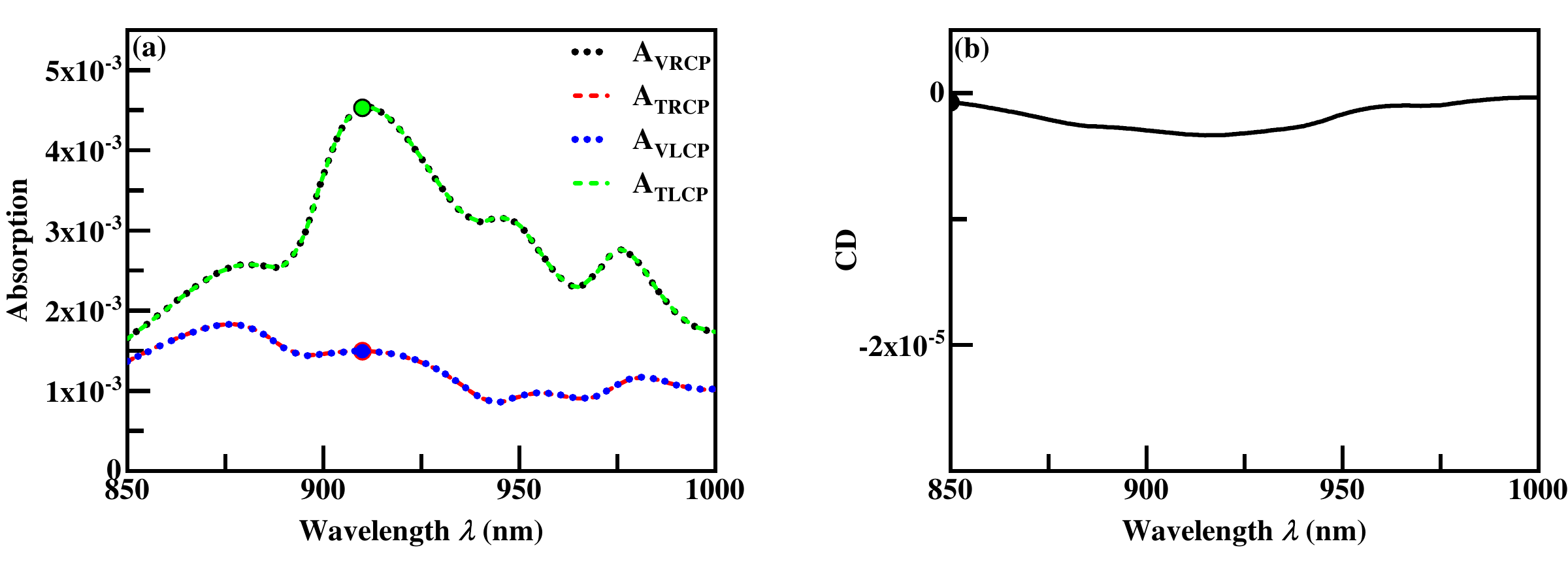}
\caption{Optical response of the symmetric CSCS with $d_t=d_l=120\,\mathrm{nm}$. (a) Absorption spectra of the transverse and longitudinal nanorods under RCP and LCP illumination. (b) CD spectrum of the complete structure.}
\label{abssy}
\end{figure}

Building on the previous findings, we further explored the chiral properties of a CSCS consisting of one transverse and one longitudinal nanorod. To facilitate comparison with the single-nanorod case, we introduced an offset between the two nanorods. This offset reduces their interaction and minimizes changes in the surface plasmon resonance modes. Specifically, both the transverse and longitudinal offsets were set to $d_t=d_l=120\,\mathrm{nm}$. Figure~\ref{abssy} presents the absorption spectra of the transverse and longitudinal nanorods under RCP and LCP illumination in panel (a), together with the CD spectrum in panel (b). A prominent absorption feature appears around $910\,\mathrm{nm}$, which was therefore chosen as the reference wavelength for further analysis of the chiroptical properties. To facilitate comparison with the CD of the chiral CSCS, an expanded vertical scale was used in Fig.~\ref{abssy}(b). Theoretically, the CD of this achiral CSCS should be zero; however, small nonzero values are observed in the calculated CD spectrum because of numerical errors.

For isolated nanorods, the oscillations of the transverse and longitudinal nanorods are identical under both RCP and LCP excitation after accounting for their orientations. However, in the coupled CSCS, the oscillations of the transverse and longitudinal nanorods differ under RCP and LCP excitation. Specifically, the oscillation of the transverse nanorod under RCP excitation differs from that of the longitudinal nanorod, and similarly, the oscillation of the longitudinal nanorod under LCP excitation differs from that of the transverse nanorod. Notably, the resonance of the transverse nanorod under LCP excitation precisely matches that of the longitudinal nanorod under RCP excitation, indicating identical oscillation patterns.

\begin{figure}[htbp]
\centering
\includegraphics[width=1\textwidth]{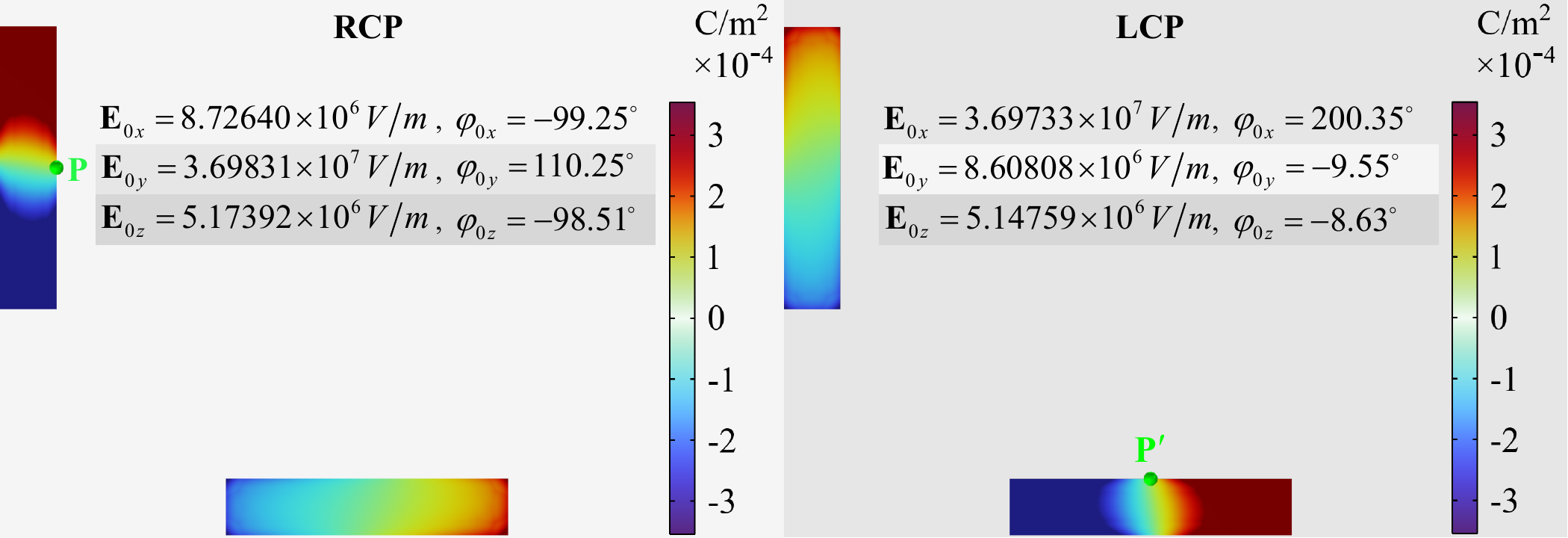}
\caption{Surface charge density distributions of the symmetric CSCS under RCP (left) and LCP (right) excitation at $910\,\mathrm{nm}$. Point \textbf{P} is located at the midpoint of the corresponding edge of the longitudinal nanorod, whereas point $\mathbf{P}^{\prime}$ is located at the midpoint of the corresponding edge of the transverse nanorod. The amplitudes and initial phases of the electric-field components in the $x$, $y$, and $z$ directions at these two points are also shown.}
\label{syshort}
\end{figure}

Figure~\ref{syshort} illustrates the charge distributions in the CSCS and clearly shows this relationship. To quantitatively demonstrate this correspondence, two symmetry-related points, \textbf{P} and $\mathbf{P}^{\prime}$, located at the midpoints of the corresponding edges of the longitudinal and transverse nanorods, respectively, were analyzed. Their oscillation behaviors were characterized by calculating the electric-field components. Figure~\ref{syshort} also presents the amplitudes and initial phases of the electric-field components in the $x$, $y$, and $z$ directions at point \textbf{P} under RCP excitation and at point $\mathbf{P}^{\prime}$ under LCP excitation. At point \textbf{P} under RCP illumination, the amplitude and phase values are $E_{0x}=8.72640\times 10^6\,\mathrm{V/m}$, $E_{0y}=3.69831\times 10^7\,\mathrm{V/m}$, and $E_{0z}=5.17392\times 10^6\,\mathrm{V/m}$, with phases $\phi_{0x}=-99.25^\circ$, $\phi_{0y}=110.25^\circ$, and $\phi_{0z}=-98.51^\circ$. At point $\mathbf{P}^{\prime}$ under LCP illumination, the corresponding values are $E_{0x}=3.69733\times 10^7\,\mathrm{V/m}$, $E_{0y}=8.60808\times 10^6\,\mathrm{V/m}$, and $E_{0z}=5.14759\times 10^6\,\mathrm{V/m}$, with phases $\phi_{0x}=200.35^\circ$, $\phi_{0y}=-9.55^\circ$, and $\phi_{0z}=-8.63^\circ$. The electric-field component along the $x$ direction in the longitudinal nanorod under RCP excitation corresponds directly to the $y$-direction component in the transverse nanorod under LCP excitation, whereas the $y$ component in the longitudinal nanorod under RCP excitation corresponds to the $x$-direction component in the transverse nanorod under LCP excitation. Furthermore, these correspondences exhibit an approximate $90^\circ$ phase shift, consistent with the expected transformation between the orthogonal nanorods under circularly polarized illumination. This analysis demonstrates that each nanorod exhibits a coupling-induced chiral response within the CSCS, although the symmetric two-nanorod structure itself remains achiral. A single nanorod exhibits this coupling-induced response because the influence of the other nanorod differs under RCP and LCP excitation. The complete CSCS remains achiral because the responses of the two nanorods are enantiomerically related and cancel in the overall response.

\begin{figure}[htbp]
\centering
\includegraphics[width=1\textwidth]{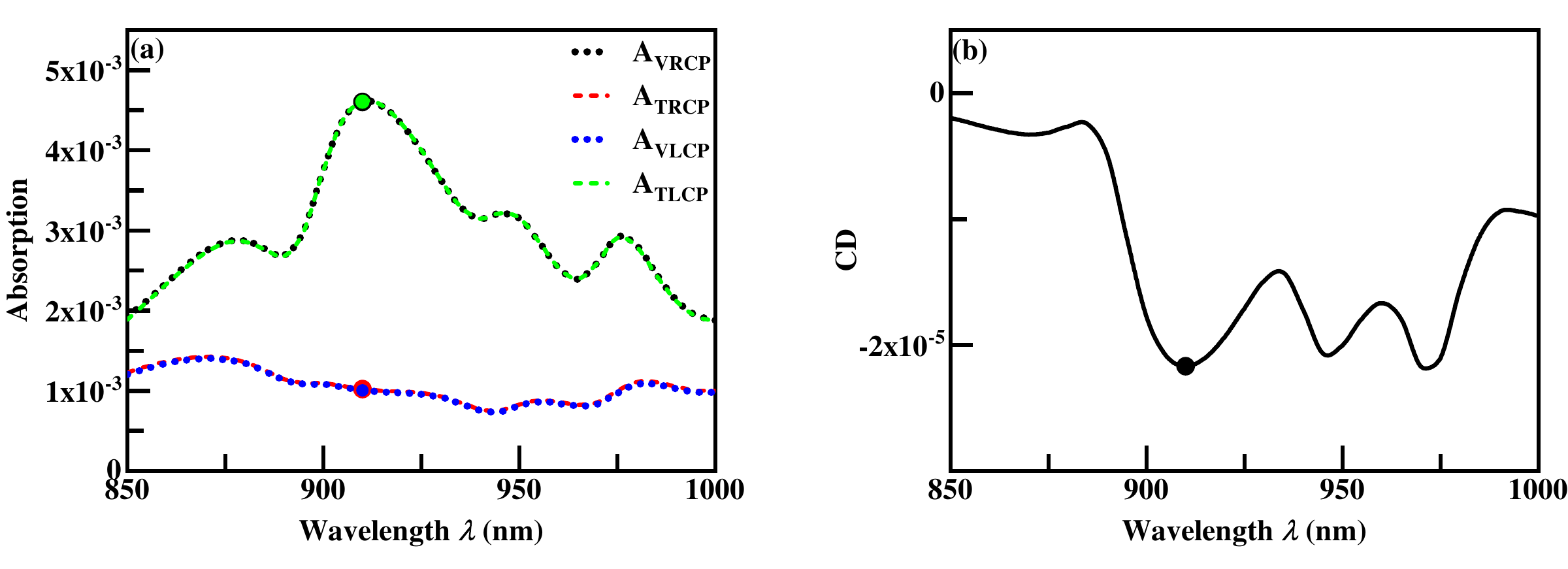}
\caption{Optical response of the asymmetric CSCS with $d_t=20\,\mathrm{nm}$ and $d_l=120\,\mathrm{nm}$. (a) Absorption spectra of the transverse and longitudinal nanorods under RCP and LCP illumination. (b) CD spectrum of the complete structure.}
\label{absassy}
\end{figure}

Finally, we investigated the relationship between the chirality-generation mechanism and the resonance modes of the chiral structure. To explore this relationship, we modified the original CSCS into a chiral structure. Specifically, we maintained the longitudinal offset at $d_l=120\,\mathrm{nm}$ while adjusting the transverse offset to $d_t=20\,\mathrm{nm}$. Figure~\ref{absassy} presents the absorption and CD spectra. The absorption spectra are similar to those of the achiral CSCS because of the relatively small structural change; however, a pronounced CD dip appears near $910\,\mathrm{nm}$.

\begin{figure}[b]
\centering
\includegraphics[width=1\textwidth]{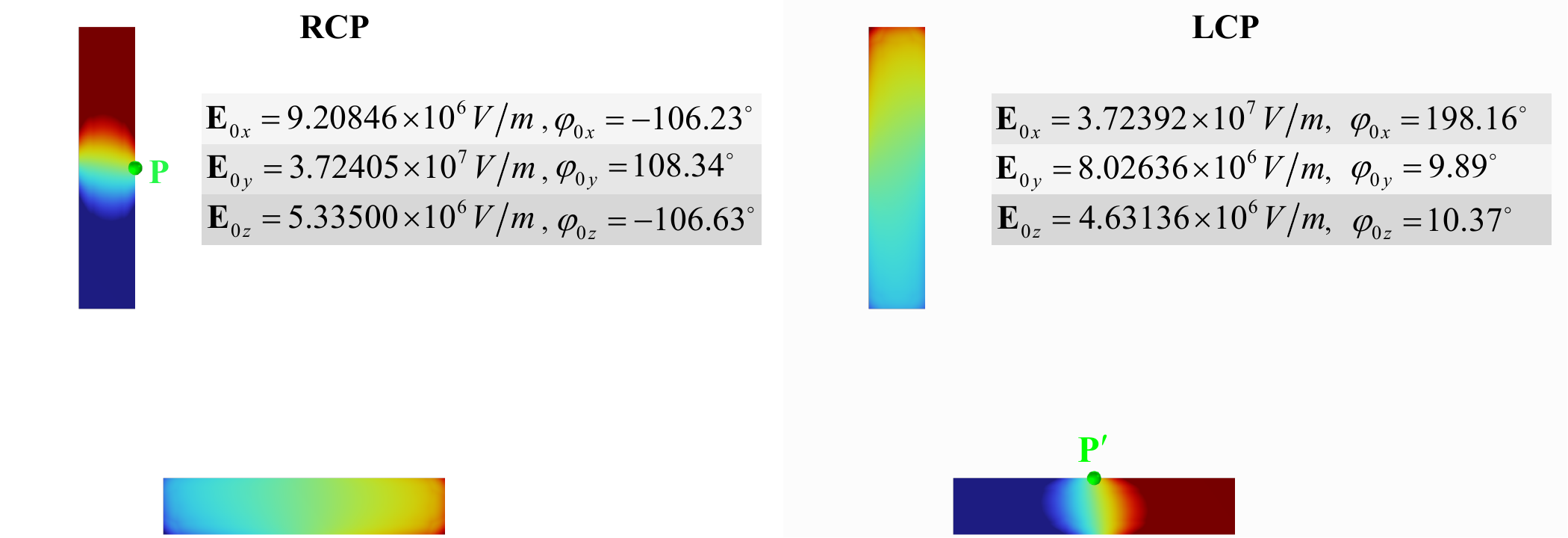}
\caption{Surface charge distributions of the asymmetric CSCS under RCP and LCP illumination at $910\,\mathrm{nm}$. The amplitudes and initial phases of the $x$, $y$, and $z$ components of the electric field are shown at point \textbf{P} under RCP excitation and at point $\mathbf{P}^{\prime}$ under LCP excitation.}
\label{assyshort}
\end{figure}

Figure~\ref{assyshort} presents the charge distributions and the amplitudes and initial phases of the electric-field components in the $x$, $y$, and $z$ directions at point \textbf{P} under RCP excitation and at point $\mathbf{P}^{\prime}$ under LCP excitation. Compared with the achiral case, the charge distribution at the initial phase changes significantly. Not only do both nanorods exhibit coupling-induced chiral responses, but the entire CSCS also becomes chiral. To quantitatively demonstrate this result, we investigated the oscillation equations at points \textbf{P} and $\mathbf{P}^{\prime}$. At point \textbf{P} under RCP illumination, the amplitude and phase values are $E_{0x}=9.20846\times 10^6\,\mathrm{V/m}$, $E_{0y}=3.72405\times 10^7\,\mathrm{V/m}$, and $E_{0z}=5.33500\times 10^6\,\mathrm{V/m}$, with phases $\phi_{0x}=-106.23^\circ$, $\phi_{0y}=108.34^\circ$, and $\phi_{0z}=-106.63^\circ$. At point $\mathbf{P}^{\prime}$ under LCP illumination, the corresponding values are $E_{0x}=3.72392\times 10^7\,\mathrm{V/m}$, $E_{0y}=8.02636\times 10^6\,\mathrm{V/m}$, and $E_{0z}=4.63136\times 10^6\,\mathrm{V/m}$, with phases $\phi_{0x}=198.16^\circ$, $\phi_{0y}=9.89^\circ$, and $\phi_{0z}=10.37^\circ$. From these data, it is evident that the amplitudes at points \textbf{P} and $\mathbf{P}^{\prime}$ differ and that the phase differences are no longer all exactly $90^\circ$. In the symmetric CSCS, these differences are relatively small and are likely attributable to numerical errors. However, in the asymmetric CSCS, some of the differences are much more pronounced. These differences were calculated by comparing the corresponding electric-field components at point \textbf{P} under RCP excitation and point $\mathbf{P}^{\prime}$ under LCP excitation. Because the nanorods are orthogonal, the $x$ component at one point was compared with the $y$ component at the other, whereas the $z$ components were compared directly. Specifically, $\Delta E_{xy}$ and $\Delta\phi_{xy}$ compare the $x$ component at point $\mathbf{P}^{\prime}$ under LCP excitation with the $y$ component at point \textbf{P} under RCP excitation, whereas $\Delta E_{yx}$ and $\Delta\phi_{yx}$ compare the reverse pair.

Table~\ref{dEdphi} presents the differences in the electric-field amplitudes and initial phases at points \textbf{P} and $\mathbf{P}^{\prime}$ for both the symmetric and asymmetric CSCS configurations. In the symmetric case, the paired amplitude differences are small, and all three phase differences are close to $90^\circ$. In the asymmetric case, the changes are component dependent. For example, $\Delta E_{yx}$ increases from $0.11832\times 10^6$ to $1.1821\times 10^6\,\mathrm{V/m}$, and $\Delta\phi_{yx}$ changes from $89.70^\circ$ to $116.12^\circ$. Similarly, $\Delta E_{zz}$ increases from $0.2633\times 10^5$ to $7.0364\times 10^5\,\mathrm{V/m}$, and $\Delta\phi_{zz}$ changes from $89.88^\circ$ to $117.00^\circ$. By contrast, $\Delta E_{xy}$ remains small, and $\Delta\phi_{xy}$ remains close to $90^\circ$. This comparison shows that asymmetric coupling does not uniformly increase every component difference; rather, it breaks the common amplitude and phase correspondence present in the symmetric CSCS, supporting the conclusion that asymmetric coupling leads to a stronger chiral response.
\begin{table}[h]
\caption{Differences in the electric-field amplitudes and initial phases at points \textbf{P} and $\mathbf{P}^{\prime}$ for the symmetric and asymmetric CSCS configurations.}
\label{dEdphi}
\centering
\begin{tabular}{ccc}
\toprule
\textbf{Case} & Symmetric CSCS & Asymmetric CSCS \\
\midrule
$\Delta E$ &
\begin{tabular}{@{}c@{}}
$\Delta E_{xy}=0.980\times 10^4\,\mathrm{V/m}$,\\
$\Delta E_{yx}=1.1832\times 10^5\,\mathrm{V/m}$,\\
$\Delta E_{zz}=0.2633\times 10^5\,\mathrm{V/m}$
\end{tabular} &
\begin{tabular}{@{}c@{}}
$\Delta E_{xy}=0.130\times 10^4\,\mathrm{V/m}$,\\
$\Delta E_{yx}=1.1821\times 10^6\,\mathrm{V/m}$,\\
$\Delta E_{zz}=0.7036\times 10^6\,\mathrm{V/m}$
\end{tabular} \\
$\Delta\phi$ &
\begin{tabular}{@{}c@{}}
$\Delta\phi_{xy}=90.10^\circ$,\\
$\Delta\phi_{yx}=89.70^\circ$,\\
$\Delta\phi_{zz}=89.88^\circ$
\end{tabular} &
\begin{tabular}{@{}c@{}}
$\Delta\phi_{xy}=89.82^\circ$,\\
$\Delta\phi_{yx}=116.12^\circ$,\\
$\Delta\phi_{zz}=117.00^\circ$
\end{tabular} \\
\bottomrule
\end{tabular}
\end{table}

According to the above results, the mechanism of chirality in the CSCS can be summarized as follows: chirality in the coupled nanorod structure arises from asymmetric resonance modes. For an isolated nanorod, the responses under RCP and LCP illumination are related by symmetry, and therefore no CD is produced. However, in the coupled configuration, the interactions between the transverse and longitudinal nanorods become asymmetric under RCP and LCP excitation.
Under RCP excitation, the transverse nanorod affects the longitudinal nanorod differently from the way it does under LCP excitation. This asymmetric coupling is amplified by resonance, leading to increased asymmetry in the electric-field and charge distributions. The resulting asymmetry in the near-field interactions between the nanorods enhances the chiral response of the system. Thus, chirality is generated by the asymmetric coupling of the nanorods, which is further amplified by resonance.

\section{Conclusion}

In this study, we explored the mechanism of chirality generation and CD in chiral nanostructures, focusing on asymmetric plasmon resonance mode distributions. By analyzing the near-field symmetry of individual nanorods and the CSCS, we demonstrated that chirality arises from asymmetric electric-field distributions between the coupled nanorods.
Our full-field calculations and quantitative analysis of the electric-field components show that the observed chiroptical effects are not caused by simple resonance shifts but by the asymmetric coupling between the transverse and longitudinal nanorods. This mechanism is generalizable and provides a broader explanation for CD generation in a variety of chiral systems, rather than being limited to the CSCS.
The absorption spectra and the pronounced CD feature near $910\,\mathrm{nm}$ support this theoretical framework and provide insights into the design of chiroptical materials.

\begin{backmatter}
\bmsection{Funding} This work was supported by the Kashi University High-Level Talent
Research Start-Up Fund (Grant No.~022024291) and the Kashi University
Key Project (Grant No.~022024572).
\bmsection{Disclosures} The authors declare no conflicts of interest.

\bmsection{Data availability} Data underlying the results presented in this paper are not publicly available at this time but may be obtained from the authors upon reasonable request.
\end{backmatter}

\bibliography{sample}
\end{document}